\documentclass{osa-article}

\journal{oe}

\begin{document}

\title{Real-time and high-throughput Raman signal extraction and processing in CARS hyperspectral imaging}

\author{Charles H. Camp Jr.\authormark{*}, John S. Bender, and Young Jong Lee}

\address{Biosystems and Biomaterials Division, National Institute of Standards and Technology, Gaithersburg, MD 20899, USA\\}

\email{\authormark{*}charles.camp@nist.gov} 



\begin{abstract*}
We present a new collection of processing techniques, collectively ``factorized Kramers--Kronig and error correction" (fKK-EC), for (a) Raman signal extraction, (b) denoising, and (c) phase- and scale-error correction in coherent anti-Stokes Raman scattering (CARS) hyperspectral imaging and spectroscopy. These new methods are orders-of-magnitude faster than conventional methods and are capable of real-time performance, owing to the unique core concept: performing all processing on a small basis vector set and using matrix/vector multiplication afterwards for direct and fast transformation of the entire dataset. Experimentally, we demonstrate that a 703\thinspace026 spectra image of chicken cartilage can be processed in 70 s ($\approx$ 0.1 ms / spectrum), which is $\approx$ 70 times faster than with the conventional workflow ($\approx $7.0 ms / spectrum). Additionally, we discuss how this method may be used for machine learning (ML) by re-using the transformed basis vector sets with new data.  Using this ML paradigm, the same tissue image was processed (post-training) in $\approx$ 33 s, which is a speed-up of $\approx$ 150 times when compared with the conventional workflow. 
\end{abstract*}

\section{Introduction}
Though long promised, coherent anti-Stokes Raman scattering (CARS)  spectroscopic microscopy (microspectroscopy) has only recently demonstrated broadband hyperspectral biological imaging at acquisition rates far in excess of what traditional Raman microspectroscopy can provide\cite{Camp2014,CampJr2015,DiNapoli2014,Pegoraro2014,Chowdary2010,Kinegawa2019}. With an imaging speed as fast as 50\thinspace000 spectra per second\cite{Tamamitsu2016}, a new fundamental challenge arises: high throughput extraction of Raman vibrational information from the raw CARS spectra.

CARS spectra are quintessentially a coherent mixture of photons generated through vibrationally resonant (Raman) and nonresonant (electronic) processes. The electronic contribution is typically referred to as the ``nonresonant background" (NRB) and is the root cause of CARS spectral distortion. Thus, a significant effort was made in the early years of CARS microscopy development to reduce the NRB via optical means\cite{Cheng2001,Garbacik2011,Dudovich2002,Potma2006}. The NRB, however, behaves as a stable homodyne amplifier for the Raman-generated signal; thus, reducing the NRB also reduces the Raman signal. So important is the NRB's role in signal amplification\cite{Muller2007}, that without it CARS may show little to no benefit over spontaneous Raman spectroscopy for biological imaging\cite{Cui2009}.

Unlike additive fluorescent background signals in Raman spectroscopy, the NRB is coherent with the co-generated Raman-resonant CARS components; thus, it may amplify weak signals above the noise floor. Furthermore, there is a fixed phase relationship between the Raman- and NRB-components. This inherent property led to the realization that computational methods could be used to extract the Raman portion of the CARS spectra using so-called ``phase-retrieval methods": the Kramers--Kronig relation (KK)\cite{Liu2009} or the maximum entropy method\cite{Vartiainen1992}. These early works assumed that the NRB was either known \textit{a priori} or the NRB of a surrogate material (e.g. coverslip glass, water, salt\cite{Karuna2016}) was appropriate. Later, it was demonstrated that using surrogate materials for NRB approximations led to amplitude and phase errors that were linked analytically\cite{Camp2016}. These errors could be corrected using ``phase-error" correction (PEC) and ``scale-error" correction (SEC) methods\cite{Camp2016}, which also reveals the relationship between the actual NRB and the surrogate. Importantly, this relationship demonstrated that CARS is unique among imaging techniques: it is inherently self-referencing. The spectral ratio of the Raman component to the actual NRB is an inherent property of a molecular system; thus, this ratio is maintained even in the case of sample scatter or absorption -- just the signal-to-noise ratio (SNR) is affected. This enables one-to-one comparison of spectra between samples and even different CARS architectures (with different laser systems and wavelengths)\cite{Camp2016}. Other coherent Raman methods, most notably stimulated Raman scattering (SRS) microscopy/spectroscopy\cite{Freudiger2008a}, do not co-generate an NRB and do not have this internal referencing ability. Thus, SRS spectra are undistorted and useful for chemical identification, but the spectral amplitudes are not necessarily directly comparable with other results, potentially challenging quantitative analysis.

To generate robust, quantitative CARS Raman spectral data and to support the rapidly increasing data rates and volumes, we have developed a series of new methods collectively referred to as ``factorized Kramers--Kronig and error correction" (fKK-EC). The new, unique principle of fKK-EC is that raw CARS spectral data can be factorized/decomposed into a small set of basis vectors on which the necessary processing steps will actually be performed. In this work, we use singular value decomposition (SVD) for its robust, accurate decomposition of matrices, although it is possible to use others as well. Previously, SVD has been used for denoising\cite{Masia2013,Camp2014,Camp2016}, but the remainder of operations were performed on the individual spectra. Additionally, matrix factorization, such as non-negative matrix factorization (NMF) / multivariate curve resolution (MCR) have been applied to post-processed data for analysis\cite{Masia2013, Zhang2013}.

The fKK-EC is composed of three parts that will be described theoretically in more detail below: phase retrieval via a factorized KK relation (fKK), factorized PEC (fPEC), and factorized SEC (fSEC). These three parts operate on the basis vectors; thus, the image data is not reconstituted between each step. This limited operation on a small number of basis vectors is economical in terms of speed and memory usage without losing the spectral information, compared with the previous methods. Furthermore, basis vector sets can be re-used on new data; thus, the fKK-EC method can be used like a machine learning method, ML:fKK-EC, for short. In this paradigm, the full fKK-EC is performed (``trained") on a portion of data (e.g., the first image), and subsequent images are able to be processed (in full) via matrix multiplication. This factorized method enables new data to be processed on-the-fly in real-time during acquisition: denoised, phase-retrieved, and phase- and scale-error corrected. Like all ML methods, this process does require that the training data reflect what will be contained in upcoming data -- though this is readily testable as will be discussed.

\section{Theory}
\subsection{Background: conventional post-processing for a single CARS spectrum}
CARS is a third-order nonlinear scattering phenomenon in which two photons (``pump" and ``Stokes") excite a Raman vibrational mode from which a third photon (``probe") inelastically scatters\cite{CampJr2015}. Furthermore, this process does not happen in isolation and other nonlinear processes, such as degenerate four-wave mixing, may occur, leading to the generation of a so-called nonresonant background (NRB). So ubiquitous is the NRB that theoretical treatments of the CARS mechanism automatically incorporate the NRB, and the term ``CARS signal" implies a coincident NRB. Thus, in this manner the CARS signal, $I_{CARS}$, may be described as\cite{Camp2014}:
\begin{align}
I_{CARS}(\omega) ~\propto~ \left | \left \{ \left [E_S(\omega)\star E_p(\omega)\right ]\chi^{(3)}(\omega) \right \} \ast E_{pr}(\omega)\right |^2 \approx \left|\tilde{C}_{st}(\omega)\right|^2 \left |\tilde{\chi}^{(3)}(\omega)\right |^2,\label{Eq:ICARS}
\end{align}
where $E_p$, $E_S$, and $E_{pr}$ are the frequency-domain ($\omega$) pump, Stokes, and probe fields, respectively; $\chi^{(3)}$ is the third-order nonlinear susceptibility, which is a summation of resonant ($\chi_r$; Raman vibrational) and nonresonant ($\chi_{nr}$; electronic) components, and $\tilde{C}_{st}$ is the system response function that incorporates such properties as laser source profiles, optical filter transmission profiles, and detector response. In the right-hand part of Eq. \ref{Eq:ICARS}, the tilde above $\chi^{(3)}$ is used to indicate that the nonlinear susceptibility is convolved with the probe\cite{Camp2016}; though, in the remainder of this manuscript, it will not be explicitly used. `$\star$' and `$\ast$' are the cross-correlation and convolution operators, respectively.

The overarching goal of phase retrieval methods is to extract Im$\{\chi^{(3)}(\omega)\}$ from $I_{CARS}(\omega)$, which is the equivalent material property probed by traditional Raman spectroscopy\cite{Tolles1977}. If $\tilde{C}_{st}(\omega)$ and $I_{NRB}(\omega)$ are quantitatively measurable/known, this goal would be achievable\cite{Liu2009}. However, this has not thus far been demonstrated. A more capable solution that also leads to the aforementioned self-referencing of CARS, is to calculate $K_{CARS}(\omega) \triangleq \chi^{(3)}(\omega)/\chi_{nr}(\omega)$. Using the KK formalism and assuming, for the moment, that the $I_{NRB}(\omega)$ of the sample itself is measurable\cite{Camp2016}:
\begin{align}
    K_{CARS}(\omega) = \frac{\chi^{(3)}(\omega)}{\chi_{nr}(\omega)}= \underbrace{\sqrt{\frac{I_{CARS}(\omega)}{I_{NRB}(\omega)}}}_{A_{CARS}(\omega)} \exp \left[i\underbrace{ \hat{\mathcal{H}}\left\{\frac{1}{2}\ln\frac{I_{CARS}(\omega)}{I_{NRB}(\omega)}\right\}}_{\phi_{CARS}(\omega)}\right],\label{Eq:KCARS}
\end{align}
where $\hat{\mathcal{H}}$ is the Hilbert transform. To approximate the NRB, one uses a surrogate/reference material with nonlinear susceptibility $\chi_{ref}(\omega)$, which leads to a CARS signal, $I_{ref}(\omega)$. One can model this relationship between the actual NRB and the surrogate as $I_{ref}(\omega) = \Xi \xi(\omega) I_{NRB}(\omega)$, where $\xi(\omega)$ is a frequency-dependent function and $\Xi$ is a constant. Both are real valued. It should be noted that these terms encompass differences in both the material properties as well as any optical system response changes (e.g., related to $\tilde{C}_{st}$). Applying this new scenario to Eq. \ref{Eq:KCARS}:
\begin{align}
    K(\omega) ~=&~ \frac{\chi^{(3)}(\omega)}{\chi_{ref}(\omega)}= \underbrace{\sqrt{\frac{I_{CARS}(\omega)}{I_{ref}(\omega)}}}_{A(\omega)} \exp \left[i\underbrace{ \hat{\mathcal{H}}\left\{\frac{1}{2}\ln\frac{I_{CARS}(\omega)}{I_{ref}(\omega)}\right\}}_{\phi(\omega)}\right]\nonumber\\
    ~=&~ A_{CARS}(\omega)\underbrace{\sqrt{\frac{1}{\Xi \xi(\omega)}}}_{A_{err}(\omega)}\exp \left [ i\phi_{CARS}(\omega) + i \underbrace{\hat{\mathcal{H}}\left\{\frac{1}{2}\ln\frac{1}{\Xi \xi(\omega)}\right\}}_{\phi_{err}(\omega)}\right ]\label{Eq:K}
\end{align}
From this equation one will notice that the use of a reference material has led to a multiplicative amplitude error and an additive phase error\cite{Camp2016}, which are themselves related by a KK relation. Thus, baseline detrending of Im$\{K(\omega)\}$ is not appropriate. There is a solution: PEC. Under the assumption of a slowly-varying $\xi(\omega)$, one may find the phase error using detrending methods, such as asymmetric least squares (ALS)\cite{Eilers2003,Eilers2005}, and remove it and the associated amplitude error (within a constant $\Xi$): $A_{err}(\omega) = 1/\sqrt{\Xi \xi(\omega)} = \exp\left[-\hat{\mathcal{H}}\{\phi_{err}(\omega)\}\right] / \Xi$. PEC does not account for and remove $\Xi$ as the Hilbert transform of a constant is 0 (i.e., $\hat{\mathcal{H}}\{\ln \Xi \xi(\omega)\} = \hat{\mathcal{H}}\{\ln \xi(\omega)\}$). Finding the constant $\Xi$ is the role of SEC\cite{Camp2016}. This may be calculated from the real part of $K(\omega)$ after PEC:
\begin{align}
    \frac{1}{\sqrt{\Xi}} = \left \langle \text{Re}\left\{K(\omega) \exp \left [ \hat{\mathcal{H}}\{\phi_{err}(\omega)\}\right] \exp\left[-i\phi_{err}(\omega) \right] \right\}\right \rangle,\label{Eq:Mean_real_part}
\end{align}
where `$\langle\dotsb\rangle$' indicates the mean over the frequency. Due to computational distortion of the numerical Hilbert transform, one usually does not simply use the mean but rather a trendline\cite{Camp2016}.

\begin{figure}[ht]
	\centering\includegraphics[width=\linewidth]{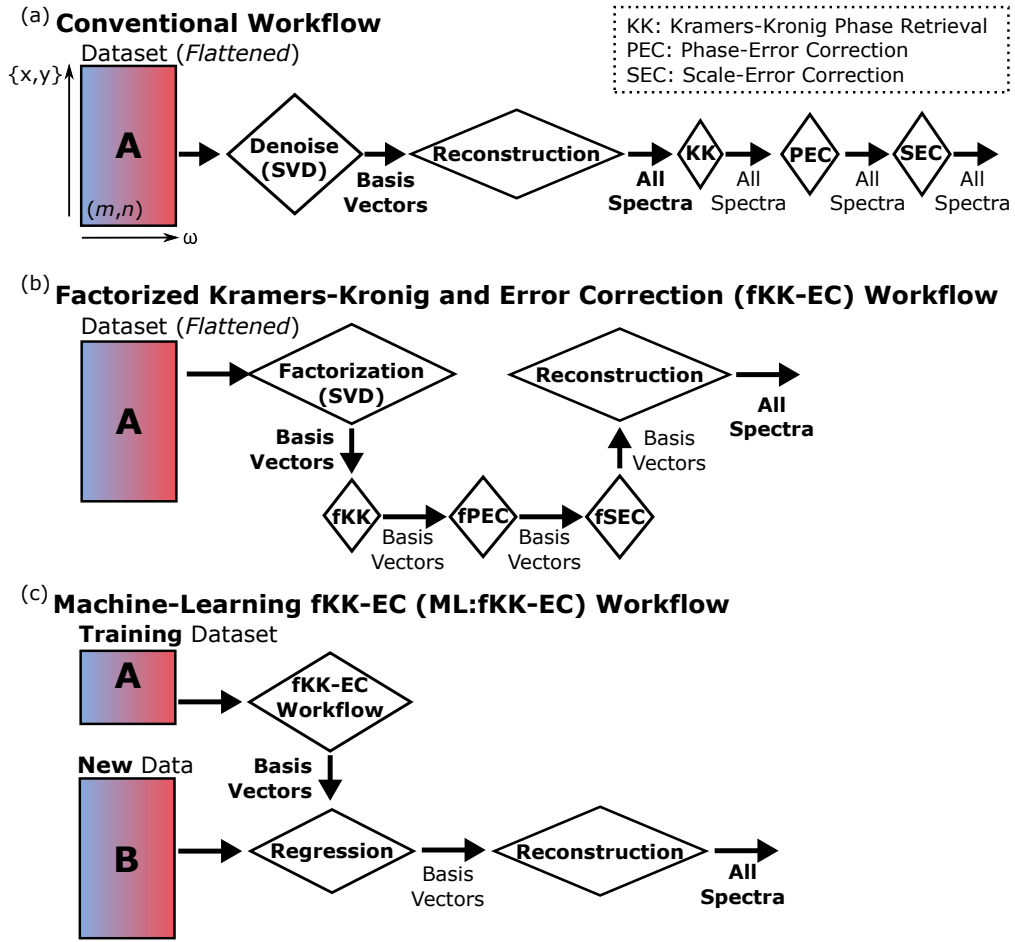}
	\caption{(a) Workflow for the conventional denoising, KK, PEC, and SEC, where $m$ is the total number of spatial pixels (``flattened") and $n$ is the number of frequency channels. (b) Workflow for the fKK-EC where the processing steps are performed on basis vectors rather than the underlying spectra. (c) Workflow for the ML:fKK-EC in which only the training data is processed via the fKK-EC and regression is used to transform new data.\label{Fig:infographics}}
\end{figure}

In summary, using the KK relation, PEC, and SEC, one can calculate $K_{CARS}(\omega)$ from $K(\omega)$ as:
\begin{align}
    K_{CARS}(\omega) = \frac{K(\omega) \exp \left [ \hat{\mathcal{H}}\{\phi_{err}(\omega)\}\right] \exp\left[-i\phi_{err}(\omega) \right]}{\left \langle \text{Re}\left\{K(\omega) \exp \left [ \hat{\mathcal{H}}\{\phi_{err}(\omega)\}\right] \exp\left[-i\phi_{err}(\omega) \right] \right\}\right \rangle}\label{Eq:KCARS_from_K}
\end{align}
Thus, without directly measuring the NRB, one can find the ratio $\chi_r(\omega)/\chi_{nr}(\omega)$ at every pixel because every pixel is self-referenced to its own nonresonant component. The full conventional workflow is illustrated in Fig. \ref{Fig:infographics}(a). This ratio is maintained even in the presence of absorption and scatter as both the Raman and NRB components are equally affected; though, the SNR deteriorates.

\subsection{SVD factorization, denoising, and fKK}
The proposed fKK-EC enables high-throughput and real-time Raman signal extraction from spectroscopic CARS data via factorization, which dramatically reduces the number of vectors for which each processing step is applied. For example, rather than independently applying to one million spectra in a one-megapixel image, the processing may be applied to 100 basis vectors. A flow chart that describes the fKK-EC workflow is shown in Figure \ref{Fig:infographics}(b).

The first step in this process is factorization of the input data. In this work, SVD decomposes a matrix $\mathbf{A}$ into three matrices as $\mathbf{A} = \mathbf{U}\mathbf{S}\mathbf{V}^H$. The $H$-superscript indicates the Hermitian transpose; $\mathbf{U}$ and $\mathbf{V}$ are unitary matrices whose columns are the left- and right singular vectors, respectively; and $\mathbf{S}$ is a diagonal matrix whose entries are known as singular values (we will denote the vector containing just the singular values as $\mathbf{s}$). In this work, we explicitly assume that the dataset is oriented so that row-number ($m$) corresponds to spatial components (see Fig. \ref{Fig:infographics}(a)) and column-number ($n$) to frequency. Thus $\mathbf{U}$ is composed of spatial basis vectors while $\mathbf{V}$, spectral basis vectors. Further, $\mathbf{A}$ is real; thus, the Hermitian transpose ($H$) is a transpose ($T$) as will be indicated in the remaining derivations. The SVD\cite{Huang1975, Tufts1982, Masia2013, Camp2014, Camp2016} is widely used for denoising by removing noise-dominant singular vectors that [ideally] only contribute to noise. This is accomplished by either setting singular values corresponding to noise-related singular vectors to 0, or equivalently creating new $\mathbf{U}$, $\mathbf{S}$, and $\mathbf{V}$ matrices that exclude the non-desired singular values and vectors, which leads to reduced data volumes. We have implemented the latter in our simulations and experiments. Note that in the remaining derivations we do not explicitly denote whether $\mathbf{U}$, $\mathbf{S}$, or $\mathbf{V}$ were altered for denoising; though, all derivations remain valid.

For the fKK, one would conceptually apply the KK relation to the spectral basis vectors in $\mathbf{V}$. However, this is not appropriate because of the log-function in the KK (Eq. \ref{Eq:KCARS}) and the orthonormal nature of SVD singular vectors (positive- and negative-values). Rather we apply the SVD to $\ln \sqrt{I^{[m]}_{CARS}(\omega)/I_{ref}(\omega)} = \mathbf{a}_m(\omega)$, where the $m$-superscript denotes the m$^{th}$ spectrum, which leads to the $\mathbf{a}_m$-vector. For the following derivation, we assume that we have $M$ spectra, and the $\omega$ vector is $N$-frequency increments long. Thus, $\mathbf{A}$ may be written as:
\begin{align}
    \mathbf{A} ~=&~ \left[\mathbf{a}_1(\omega), ~\mathbf{a}_2(\omega), ~\dots~, ~\mathbf{a}_M(\omega)\right ]^T \in \mathbb{R}^{M\times N}= \mathbf{U}\mathbf{S}\mathbf{V}^T
\end{align}
Assuming a "reduced" SVD implementation, we have more spectra than the length of the frequency vector $\omega$ (i.e., M $>$ N); thus, $\mathbf{U} \in \mathbb{R}^{M,N}$, $\mathbf{s} \in \mathbb{R}^N; \text{diag}(\mathbf{s}) = \mathbf{S} \in \mathbb{R}^{N,N}$, and $\mathbf{V} \in \mathbb{R}^{N,N}$. As the $\mathbf{U}$- and $\mathbf{S}$-elements act as constant weighting terms to the right singular vectors ($\mathbf{V}$'s columns) and the Hilbert transform is a linear operator\cite{Poularikas1999}, this is equivalent to only applying the transform to the right singular vectors:
\begin{align}
    \hat{\mathcal{H}}\{\mathbf{A}\} ~=&~ \begin{bmatrix}
           \hat{\mathcal{H}}\{A_{1,:}(\omega)\}\\
           \hat{\mathcal{H}}\{A_{2,:}(\omega)\}\\
           \vdots \\
           \hat{\mathcal{H}}\{A_{M,:}(\omega)\}
         \end{bmatrix} = 
         \begin{bmatrix}
           \hat{\mathcal{H}}\{U_{1,1}S_{1,1}\mathbf{v}_1(\omega) + U_{1,2}S_{2,2}\mathbf{v}_2(\omega) + \dots + U_{1,N}S_{N,N}\mathbf{v}_N(\omega)\} \\
           \hat{\mathcal{H}}\{U_{2,1}S_{1,1}\mathbf{v}_1(\omega) + U_{2,2}S_{2,2}\mathbf{v}_2(\omega) + \dots + U_{2,N}S_{N,N}\mathbf{v}_N(\omega)\} \\
           \vdots \\
           \hat{\mathcal{H}}\{U_{M,1}S_{1,1}\mathbf{v}_1(\omega) + U_{M,2}S_{2,2}\mathbf{v}_2(\omega) + \dots + U_{M,N}S_{N,N}\mathbf{v}_N(\omega)\}
         \end{bmatrix}\nonumber\\
         ~=&~ \begin{bmatrix}
           U_{1,1}S_{1,1}\hat{\mathcal{H}}\{\mathbf{v}_1(\omega)\} + U_{1,2}S_{2,2}\hat{\mathcal{H}}\{\mathbf{v}_2(\omega)\} + \dots + U_{1,N}S_{N,N}\hat{\mathcal{H}}\{\mathbf{v}_N(\omega)\} \\
           U_{2,1}S_{1,1}\hat{\mathcal{H}}\{\mathbf{v}_1(\omega)\} + U_{2,2}S_{2,2}\hat{\mathcal{H}}\{\mathbf{v}_2(\omega)\} + \dots + U_{2,N}S_{N,N}\hat{\mathcal{H}}\{\mathbf{v}_N(\omega)\} \\
           \vdots \\
           \hat{\mathcal{H}}\{U_{M,1}S_{1,1}\mathbf{v}_1(\omega)\} + U_{M,2}S_{2,2}\hat{\mathcal{H}}\{\mathbf{v}_2(\omega)\} + \dots + U_{M,N}S_{N,N}\hat{\mathcal{H}}\{\mathbf{v}_N(\omega)\}
         \end{bmatrix}\nonumber\\
         ~=&~ \mathbf{U}\mathbf{S}\hat{\mathcal{H}}\{\mathbf{V}^T\}\label{Eq:HilbertA_Derivation}
\end{align}
The total fKK process without PEC or SEC may be described as:
\begin{align}
    \mathbf{K}_{fKK}(\omega) = \exp \left (\mathbf{A}\right) \exp \left (i \mathbf{U}\mathbf{S}\hat{\mathcal{H}}\{\mathbf{V}^T\}\right)\label{Eq:fKK}
\end{align}

As an addendum to this derivation, we will discuss considerations under the case of mixed Poisson-Gaussian noise (heteroscedastic noise generally). In previous work\cite{Camp2016}, denoising was improved via the use of an Anscombe transformation prior to SVD. As Poisson noise is not additive\cite{Giryes2014}, SVD is often impaired in separating signal and noise. The Anscombe transform aims to convert a signal with mixed noise into a signal with unit variance. Though advantageous, this nonlinear transform is not compatible with the current fKK derivation. Thus, to improve denoising, there are 2 options: (1) denoise before the fKK using the Anscombe transformation and SVD (then reconstruction), or (2) apply a scaling term $f(\omega)$ to $A(\omega)$, which is the same for each spectrum. In simulations and experiments below, we apply the latter. The scaling term we selected was inspired by the purpose of the Anscombe transformation: normalizing variance. Suppose we have an image of all the same spectrum, the standard deviation ($\sigma_\mathbf{A}(\omega)$) of the previously defined $\mathbf{A}$ may be approximated as\cite{Ku1966}: 
\begin{align}
    \sigma_\mathbf{A}(\omega) ~\approx&~ \frac{\sigma_{CARS}(\omega)}{2\langle I_{CARS}\rangle(\omega)} \approx \frac{\sqrt{\alpha \langle I_{CARS}\rangle(\omega) + \sigma_g^2}}{2 \langle I_{CARS}\rangle(\omega)} \triangleq \frac{1}{f(\omega)},
\end{align}
where $\langle\dotsb\rangle$ indicates the mean spectrum, $\alpha$ is a Poisson noise multiplier, and $\sigma_g$ is the standard deviation of the additive white Gaussian noise. We have assumed that the $I_{ref}(\omega)$ used is effectively noiseless as the reference spectra is often an averaged and/or denoised version of repeated measurements of a surrogate material. Applying this scaling term, the fKK would be re-written as:
\begin{align}
    A_{m,:} ~=&~ f(\omega) \frac{1}{2}\ln\frac{I^{[m]}_{CARS}(\omega)}{I_{ref}(\omega)}\quad\text{for all } m\\
    \mathbf{K}_{fKK}(\omega) ~=&~ \exp \left ( \frac{\mathbf{A}}{f(\omega)}\right )\exp \left (i \mathbf{U} \mathbf{S} \hat{\mathcal{H}}\left\{\frac{\mathbf{V}^T}{f(\omega)}\right\}\right )
\end{align}
where $A_{m,:}$ is the m$^{th}$ spectrum (row) in $\mathbf{A}$. 

In the remainder of this manuscript, we will include $f(\omega)$ in the derivations; though, this factor can be set to one in the case of pure additive white Gaussian noise. Mathematical notation note: we are explicitly writing $f(\omega)$ to emphasize that it is a single spectrum, and when it divides a matrix, it is applied along the spectral axis (e.g., each row of $\mathbf{A}$ or $\mathbf{V}^T$).

\subsection{Factorized PEC (fPEC)}
PEC is the process of finding the phase error caused by using a surrogate reference material as an approximation for the sample NRB. In the factorized context:
\begin{align}
     \mathbf{\Phi}_{err} ~=&~ \mathcal{D}\left\{\mathbf{U} \mathbf{S}  \hat{\mathcal{H}}\left\{\frac{\mathbf{V^T}}{f(\omega)}\right\}\right\}\approx \mathbf{U} \mathbf{S} \mathbf{\Phi}^T_{PEC}
\end{align}
where $\mathcal{D}$ is a detrending operator, and $\mathbf{\Phi}_{PEC}$ is a basis set describing phase error. We do not want to detrend every spectrum as described in the proceeding equation and the orthonormal $\mathbf{V}$ singular vectors are not readily usable for baseline detrending as they often have positive and negative values with no clear baseline. Rather we will take the approach of sub-sampling $\mathbf{U}$ (to form $\mathbf{U}_{ss}$), calculate $\mathbf{\Phi}_{err}$, and regress to approximate $\mathbf{\Phi}_{PEC}$. This dramatically reduces the computational burden compared to using the full $\mathbf{U}$. Our current practice, inspired by vertex component analysis (VCA)\cite{Nascimento2005}, is to sub-sample $\mathbf{U}$ by keeping the rows of $\mathbf{U}$ that have the highest and lowest values for each column:, and optionally a sub-sample between. For the maximum and minimum:
\begin{align}
    \mathbf{q}_{max} ~=&~ \text{argmax}_i\{\mathbf{U}_{:,i}\}\quad\text{for each }i\\
    \mathbf{q}_{min} ~=&~ \text{argmin}_i\{\mathbf{U}_{:,i}\}\quad\text{for each }i\\
    \mathbf{U}_{ss} ~=&~ \mathbf{U}_{\mathbf{q},:}
\end{align}
where the `:' indicates all row or column entries, and $\textbf{q} = \mathbf{q}_{min}~\cup~\mathbf{q}_{max}$ indicates the union row indices. $\mathbf{q}$ can also contain a sub-sample between the max- and min-values for each column $\mathbf{U}$. From this:
\begin{align}
    \mathbf{\Phi}_{err} ~=&~ \mathcal{D}\left\{\mathbf{U}_{ss} \mathbf{S}  \hat{\mathcal{H}}\left\{\frac{\mathbf{V^T}}{f(\omega)}\right\}\right\}\approx \underbrace{\mathbf{U}_{ss} \mathbf{S}}_{\mathbf{X}} \mathbf{\Phi}^T_{PEC}\label{Eq:phi_err_USphi_b}\\
    \mathbf{\Phi}^T_{PEC} ~=&~ \mathbf{X}^{-1}\mathbf{\Phi}_{err}\Longrightarrow \mathbf{\Phi}^T_{PEC} ~=~
    \left (\mathbf{X}^T\mathbf{X}  + \lambda \mathbf{I}\right)^{-1}\mathbf{X}^T\mathbf{\Phi}_{err}\label{Eq:PEC_Ridge}
\end{align}
where $\lambda$ is a non-negative scalar regularization weight and $\mathbf{I}$ is an identity matrix. The left-hand statement in Eq. \ref{Eq:PEC_Ridge} is an ordinary least-squares regression using a [pseudo]-inverse. In practice, however, this result is unstable owing to significant multicollinearity in the singular vectors. These collinearities cause erroneously large $\mathbf{\Phi}_{PEC}$ entries, especially those corresponding to the smallest singular values. One solution to this problem is ridge regression (also known as Tikhonov regularization) as shown on the right side of Eq. \ref{Eq:PEC_Ridge}.

The action of the combined fKK and fPEC without fSEC can be described as:
\begin{align}
    \mathbf{K}_{fKK-fPEC}(\omega) ~=&~ \exp \left [ \mathbf{U}\mathbf{S}\left (\frac{\mathbf{V}^T}{f(\omega)} + \hat{\mathcal{H}}\{\mathbf{\Phi}_{PEC}\}\right)\right ]\exp \left [i \mathbf{U} \mathbf{S}\left( \hat{\mathcal{H}}\left\{\frac{\mathbf{V}^T}{f(\omega)}\right\} - \mathbf{\Phi}_{PEC}\right)\right ]\label{Eq:fPEC_K}
\end{align}
noting that the amplitude and phase terms are still related by a Hilbert transform.

\subsection{Factorized SEC (fSEC)}
In the conventional form of the SEC, the PEC-corrected spectra are divided by the mean of the real part as described in Eq. \ref{Eq:KCARS_from_K}. An alternative and equivalent approach is to calculate the mean of the natural log of the magnitude of the PEC-corrected spectra:
\begin{align}
    \left \langle \frac{1}{2}\ln \frac{I_{CARS}(\omega)}{I_{NRB}(\omega)\Xi} \right \rangle ~=&~ \left \langle \frac{1}{2} \ln \frac{I_{CARS}(\omega)}{I_{NRB}(\omega)} \right \rangle - \frac{1}{2}\ln \Xi = \ln \frac{1}{\sqrt{\Xi}}\label{Eq:Mean_PEC_Corrected_Magnitude}
\end{align}
It should be noted that the mean of the first expression in the previous equation can be solved analytically, for example, using partial fraction decomposition, assuming that $\chi_r(\omega) = \sum_m A_m / (\Omega_m - \omega -i\Gamma_m)$, $A_m$, $\Omega_m$, and $\Gamma_m$ are real and positive-valued, and $\chi_{nr}$ is constant and positive, real-valued.

The left-hand expression in Eq. \ref{Eq:Mean_PEC_Corrected_Magnitude} for the dataset is equivalent to the magnitude of the term inside the exponential function in Eq. \ref{Eq:fPEC_K} as:
\begin{align}
    \left \langle \mathbf{U} \mathbf{S} \left (\frac{\mathbf{V}^T}{f(\omega)} + \hat{\mathcal{H}}\{\mathbf{\Phi}_{PEC}\}\right)\right \rangle ~=&~ \ln \frac{1}{\sqrt{\mathbf{\Xi}}}
\end{align}
where $\mathbf{\Xi} \in \mathbb{R}^M$ is a vector of constants.

For the fSEC, we want to avoid calculating the mean for each spectrum and to operate on the PEC-corrected right singular vector. Thus, we will incorporate an fSEC correction matrix $\mathbf{V}^T_{SEC}$ into the previous expression:
\begin{align}
    \left \langle \mathbf{U} \mathbf{S} \left (\frac{\mathbf{V}^T}{f(\omega)} + \hat{\mathcal{H}}\{\mathbf{\Phi}_{PEC}\} - \mathbf{V}^T_{SEC}\right)\right \rangle ~=&~ \ln \mathbf{1} = \mathbf{0}\label{Eq:fSEC}
\end{align}
A solution for this matrix is the subtraction of the mean of the PEC-corrected right singular vector: $\mathbf{V}^T_{SEC} = \langle \mathbf{V}^T/f(\omega) + \hat{\mathcal{H}}\{\mathbf{\Phi}_{PEC}\} \rangle$. Thus, if the mean of each corrected right singular vector is zero, the mean of the magnitude will also be zero. As we previously mentioned, due to numerical errors in the Hilbert transform, rather than a strict mean, we use a trendline function, which was previously implemented as a large-window, small-order Savitzky--Golay filter\cite{Camp2016}.
Thus:
\begin{align}
    \mathbf{V}^T_{SEC} ~=&~ \mathcal{M}\left\{\frac{\mathbf{V}^T}{f(\omega)} + \hat{\mathcal{H}}\{\mathbf{\Phi}_{PEC}\}\right\}\label{Eq:fSEC_VT}
\end{align}
where $\mathcal{M}$ is a trendline (or mean function). 

\subsection{Reconstruction and the full fKK-EC}
Using the previous descriptions of the fKK, fPEC, and fSEC, we can assemble the full fKK-EC workflow and reconstruct an approximate $\mathbf{K}_{CARS}$ (akin to Eq. \ref{Eq:KCARS_from_K} for the conventional implementation). Applying Eqs. \ref{Eq:fKK}, \ref{Eq:fPEC_K}, and \ref{Eq:fSEC_VT}:
\begin{align}
    \mathbf{K}_{CARS} ~\cong&~ \exp\left[ \mathbf{U}\mathbf{S}\left(\frac{\mathbf{V}^T}{f(\omega)} + \hat{\mathcal{H}}\{\mathbf{\Phi}_{PEC}\} - \mathbf{V}^T_{SEC}\right)\right]\exp\left[i\mathbf{U}\mathbf{S}\left(\hat{\mathcal{H}}\left\{\frac{\mathbf{V}^T}{f(\omega)}\right\} - \mathbf{\Phi}_{PEC}\right)\right]\label{Eq:fKKEC}
\end{align}
again noting that $\mathbf{U}$, $\mathbf{S}$, and $\mathbf{V}$ may be reduced in size from the original SVD for the purposes of denoising.

\subsection{The machine learning (ML) paradigm ML:fKK-EC}
As previously described, the fKK-EC methods enable high-throughput analysis at significantly higher rates than the coventional workflow. Another significant benefit of the fKK-EC methods is that they can be trained as a machine learning (ML) model, i.e., the fKK, fPEC, and fSEC are fully applied to a sub-set of data, and new data is simply projected onto the derived basis vectors (as schematically described in Fig. \ref{Fig:infographics}(c)). That is to say that new data can be transformed into denoised-Raman-retrieved (fKK, fPEC, fSEC) without explicitly applying these methods, but rather with simple matrix multiplication. We will call this workflow ``ML:fKK-EC".

Hypothetically, we are going to collect many images of a sample. We will apply the full fKK-EC method to the first (or first few) images (i.e., ``training"). This provides us with: $f(\omega)$, $\mathbf{U}$, $\mathbf{S}$, $\mathbf{V}$, $\mathbf{\Phi}_{PEC}$, and $\mathbf{V}^T_{SEC}$. One assumes that upcoming images will comprise the same chemical content (but in differing concentrations and mixture profiles). In the ML:fKK-EC method, we will not re-derive the SVD, but rather regress a new left singular vector matrix $\mathbf{U}_{new}$ (which describes the spatial mixtures of $\mathbf{S}\mathbf{V}^T$). From Eq. (7) for the new data, $\mathbf{A}_{new}$ that incorporates $f(\omega)$ as well, and solving for $\mathbf{U}_{new}$ applying ridge regression:
\begin{align}
    \mathbf{A}_{new} ~=&~ \mathbf{U}_{new}\mathbf{S}\mathbf{V}^T\\
    \mathbf{U}_{new} ~=&~ \mathbf{A}_{new}\left(\underbrace{\mathbf{S}\mathbf{V}^T}_{\mathbf{X}}\right)^{-1}\Longrightarrow \mathbf{U}_{new} ~=~ \mathbf{A}_{new}\left(\mathbf{X}^T\mathbf{X} + \lambda\mathbf{I}\right)^{-1}\mathbf{X}^T\label{Eq:ML_Unew}
\end{align}
Now, one can simply apply the $\mathbf{U}_{new}$ to Eq. \ref{Eq:fKKEC}.

The ML:fKK-EC method, as will be demonstrated in simulation and experiment below, is extremely fast. Firstly, the time-consumption of the individual steps is limited to a training dataset that is much smaller than the full dataset. Secondly, new data does not need to be subjected to the fKK, fPEC, or fSEC, but rather is converted through a series of matrix multiplications: solving for Eq. \ref{Eq:ML_Unew} and applying to Eq. \ref{Eq:fKKEC}, where all the other matrices were calculated during training. For example, on the broadband CARS (BCARS) system used to collect data for this paper, spectra require $\sim$5 ms to record, but applying the ML:fKK-EC to a new spectrum requires 10's of microseconds; thus, it can be applied to new data as it is acquired, as opposed to after all data is acquired. This advancement in CARS  microscopy affords many new opportunities not previously available, such as on-the-fly evaluation of imaging quality and rapid identification of regions-of-interest and chemical constituents.

\section{Materials and methods}
\subsection{Broadband CARS (BCARS) imaging platform and software}
Images were collected on an in-house-developed BCARS microscope that is described in detail elsewhere\cite{Camp2014}. The picosecond probe laser and femtosecond supercontinuum were 13 mW and 7.1 mW on-sample. The CCD integration time was set to 3.5 ms, which corresponds to 5 ms per pixel owing to data transfer time, stage movement, and CCD cleaning time.

The BCARS system was controlled by custom LabView software written in-house. Data files were processed in Python using NumPy, SciPy, scikit-learn, and the open-source CRIkit2 software package for Python (https://github.com/CCampJr/CRIkit2). Processing was performed on a Dell Precision 7730 laptop with a 6-core i7-8850H processor at 2.6 GHz and 64 GB of RAM.

\subsection{Chicken tissue preparation}
Chicken legs were procured from a local grocer. Hyaline cartilage tissue was harvested from the knee joint above the tibia using a scalpel. The resected tissue varied in thickness from approximately 20 $\mu$m to 40 $\mu$m, as measured by BCARS imaging (``XZ" images).

\subsection{Simulation software}
The simulations were written in Python and performed from within a Jupyter Notebook. The NumPy, SciPy, scikit-learn, Pandas, Seaborn, and CRIKit2 software packages for Python were used for processing and visualization. Simulation software will be furnished upon request and will be available in a forthcoming open-source software package for Python. The simulations were performed on the same laptop as the image processing described above.

\section{Results}
Below we present simulations and experiments to demonstrate the enhanced performance (throughput) of fKK-EC and the comparability of its results with the conventional workflow. Additionally, within the experimental results, we demonstrate the application and results from the ML:fKK-EC.

\subsection{Simulation}
\begin{figure}[htbp]
\centering\includegraphics[width=\linewidth]{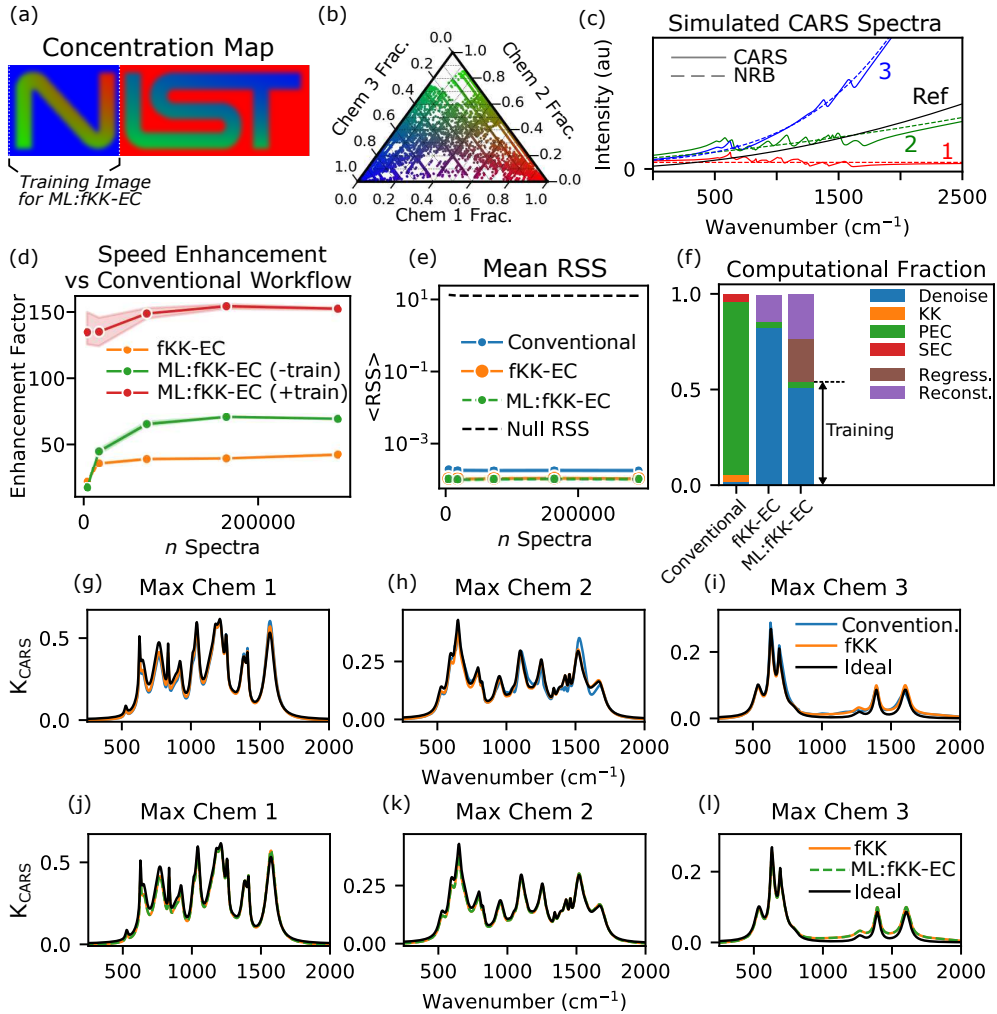}   
\caption{(a) Concentration map of simulated dataset composed of three chemical constituents, colored as red (Chem 1), green (Chem 2), and blue (Chem 3). (b) Ternary plot showing the concentration of the simulation. (c) CARS (solid lines) and NRB (dashed) spectra of the three pure constituents. The spectra of a reference surrogate is shown in black. (d) Processing speed enhancement of fKK-EC and ML:fKK-EC with respect to conventional processing. Each trace shows the mean enhancement over three runs with the shading showing $\pm$1 standard deviation. (e) Mean RSS showing the factorized methods show relatively similar, if not improved, RSS values from the conventional workflow. (f) Fraction of computing time of each step. Note: only fKK-EC and ML:fKK-EC have a reconstruction (Reconst.) step. Also, only the ML:fKK-EC has a regression (Regress.) step. (g)--(i) Comparison of single-pixel spectra processed using the conventional methods and the fKK-EC. (j)--(l) Comparison of single-pixel spectra processed using the fKK-EC and the ML:fKK-EC.\label{Fig:Sim}}
\end{figure}

We simulated a noiseless 3-chemical mixture with the concentration map shown in Figure \ref{Fig:Sim} (a) and a ternary plot of concentrations shown in \ref{Fig:Sim} (b). Chemical 1, 2, and 3 are displayed in red, green, and blue, respectively. The base dataset is 74 pixels x 246 pixels (18\thinspace204 total spectra). To analyze the fKK-EC performance versus number of spectra, this dataset is side-scaled by a factor of 0.5, 1, 2, 3, and 4; for a total of 4\thinspace551, 18\thinspace204, 72\thinspace816, 163\thinspace836, and 291\thinspace264 spectra, respectively. Synthetic Raman spectra were  generated using a summation of complex Lorentzian functions with number of peaks, amplitude, central frequency, and width being selected stochastically. Further, the real-valued $\chi_{nr}(\omega)$'s were quadratic polynomials with randomly generated non-negative coefficients, and $\chi_{ref}(\omega)$ from a linear polynomial. This approach was not chosen because of its physical realism, but rather to challenge the method \textemdash{} especially the detrending algorithm. The random number generator seed was fixed across experiments so that the same random spectra were generated. The simulated CARS spectra (and NRB) are shown in Fig. \ref{Fig:Sim}(c). The chemical spectra contain 22, 25, and 10 peaks, respectively. The spectral range of simulation was $-$500 cm$^{-1}$ to 2\thinspace500 cm$^{-1}$ sampled 810 times; though, Raman peaks could only be assigned between 500 cm$^{-1}$ to 1700 cm$^{-1}$. The stimulation profile $\tilde{C}_{st}(\omega)$ in Eq. \ref{Eq:ICARS} was set to a constant for simplicity.

Figure \ref{Fig:Sim}(d) shows the speed enhancement of the factorized methods relative to the conventional workflow. For all methods, the number of kept singular values/vectors was determined by the singular values larger than ($\max{\mathbf{A}} \times \max(M,N) \times \epsilon$), where $M$ and $N$ are the row and column dimensions of the SVD-input matrix $\mathbf{A}$, and $\epsilon$ is the "machine epsilon" for the given data type. This is the same cutoff used to estimate rank in NumPy and MATLAB software. For comparison, the time per spectrum for the conventional workflow was approximately: $\leq$100 $\mu$s for SVD and selecting basis vectors, $\leq$140 $\mu$s for the KK, $\leq$3.2 ms for the PEC, and $\leq$140 $\mu$s for the SEC; for a total of $\leq$3.6 ms / spectrum. In each conventional-method simulation run, 6 basis vectors were kept per the previously described cutoff threshold. In all factorized-method simulation runs 35~to 50~singular vectors were kept, depending on the image size. For the fKK-EC, the enhancement was $\geq$40 for all but the smallest dataset. For the 291\thinspace264 spectra simulation, for example, the total time was $<$25 seconds for all 3~replicate simulations (86 $\mu$s / spectrum). The most significant difference is the time to perform phase retrieval, with the conventional KK requiring $\approx$ 40 s and the new fKK $\approx$ 4.3 ms \textemdash{} an over 9000$\times$ improvement. The fPEC was over 1250$\times$ faster than the PEC, and the fSEC was over 3150$\times$ faster than the SEC. For the factorized workflow, the reconstruction step only added 3.3 s. Fig. \ref{Fig:Sim}(f) gives a graphical representation of the fraction of total computational time for each method. Of course it should be noted that for the ML:fKK-EC, the training fraction will reduce as more non-training data is processed.

We also compared the spectra obtained by the fKK-EC method with that of the conventional method. To that end, we calculated the residual sum-of-squares (RSS) between the extracted Raman-to-NRB ratio spectra ($K_{CARS}$ in Eq. \ref{Eq:KCARS_from_K} or Eq. \ref{Eq:fKKEC}) and the known Im$\{\chi/\chi_{nr}\}$ at each pixel. The mean RSS, $\langle$RSS$\rangle$, is shown in Fig. $\ref{Fig:Sim}$(e). For reference, the RSS if $K_{CARS}(\omega)~=~0$ (``Null RSS") is also shown. One can see that the fKK-EC and conventional workflow return similar results, with the fKK-EC being slightly better (lower). Whether this is intrinsic or due to imperfect hyperparameter tunings for each processing step (e.g., ALS parameters) will be investigated in the future as the current goal was to demonstrate approximately equivalent results. Figs. \ref{Fig:Sim}(d)-(f) compare the spectra retrieved by the conventional method and the new fKK-EC (versus the ideal) at the pixels with the maximum concentration of each simulated chemical species. In each instance, the fKK-EC spectrum returns a result closer to the ideal than the conventional method. It was determined that all errors were due to the phase error-correcting steps: the ALS could closely but not perfectly retrieve the phase error. Under a separate simulation using constant-valued NRB's, the ideal, conventional workflow, and fKK-EC all agreed ($\langle$RSS$\rangle$ $<$10$^{-14}$).

Next, we performed the same comparisons using the ML:fKK-EC implementation. The training portion of the dataset is identified in Fig. \ref{Fig:Sim}(a). Fig. \ref{Fig:Sim}(d) shows the speed enhancement of ML:fKK-EC versus the conventional workflow, both including (``+train") and excluding (``$-$train") the time used for the training portion. Thus, for a trained ML:fKK-EC system, we calculate an $\approx$150$\times$ speedup, which was $<$30 $\mu$s per spectrum for all dataset sizes. Thus, this could be performed in real-time as the data is acquired. Fig. \ref{Fig:Sim}(e) shows that the machine learning implementation provides equivalent RSS to the non-ML fKK-EC method. Fig. \ref{Fig:Sim}(f) shows the computational fraction of each step. Finally, Fig. \ref{Fig:Sim}(g)-(i) compare the retrieved from the ML:fKK-EC and non-ML version: the results are indistinguishable. 

\subsection{Experimental: chicken cartilage tissue imaging}
\begin{figure}[htbp]
	\centering\includegraphics[]{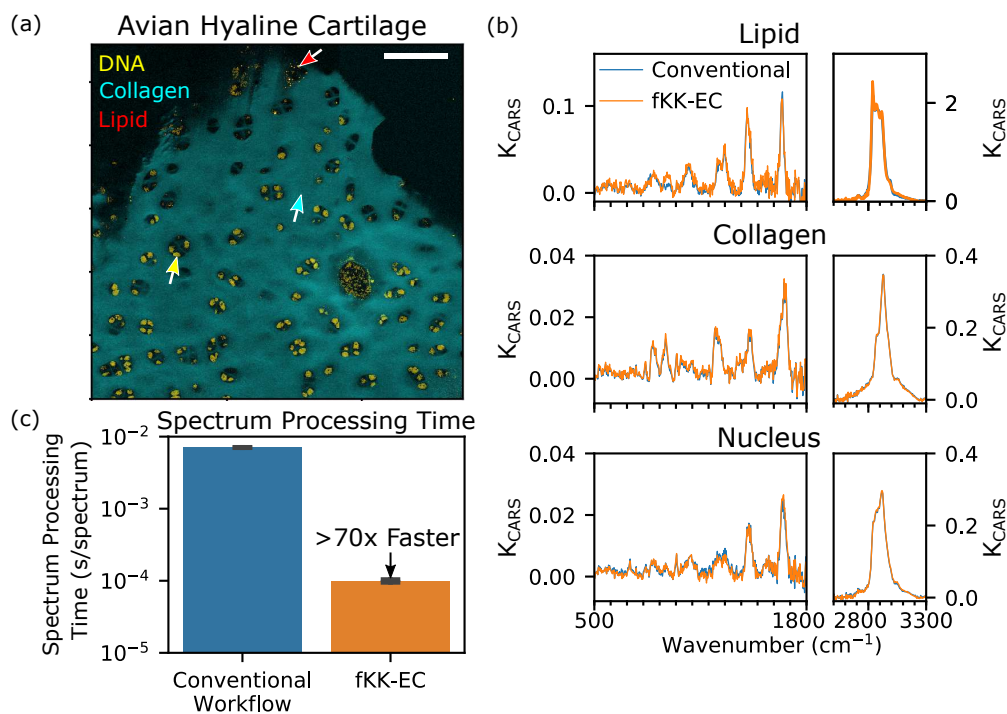}   
	\caption{(a) Pseudocolor image derived from fKK-EC processed CARS image, highlighting DNA (yellow), collagen (cyan), and lipids (red). Scale bar is 100 $\mu$m. (b) Single pixel spectra for locations identified by arrows in (a). (c) Comparison of spectrum processing time between conventional and fKK-EC workflow. \label{Fig:Chicken1}}
\end{figure}

Next, we analyzed a stitched series of BCARS images (9) of hyaline cartilage excised from chicken knee tissue. The individual original images are 300 pixels x 300 pixels, with $\approx$3\% overlap (per side) with neighboring images. The stitched image is 846 pixels by 831 pixels (703\thinspace026~pixels total). Fig. \ref{Fig:Chicken1}(a) shows a pseudocolor image from the fKK-EC process, colorizing DNA, collagen, and lipids. The DNA was highlighted utilizing the peak at 720 cm$^{-1}$. To maximize contrast between DNA and other chemical components, we used the side of this peak 716 cm$^{-1}$, subtracting a linear interpolated baseline between (691 to 738) cm$^{-1}$. Tentatively, we assign this peak to the nucleotide adenine\cite{DeGelder2007}. We did not see a strong peak at 785 cm$^{-1}$, which corresponds, in part, to phosphodiester stretch of the DNA backbone; thus, we hypothesize, that DNA-nucleases may have degraded the DNA as this is not fresh chicken tissue, but rather grocery store procured. The collagen was highlighted by 855 cm$^{-1}$ (proline ring C-C-stretch\cite{Frushour1975}) peak relative to the trough at 900 cm$^{-1}$. Lipids were highlighted using the intensity at 2837 cm$^{-1}$ (CH$_2$-symmetric stretch\cite{Czamara2015}).

Spectra retrieved using the conventional method and the fKK-EC are shown in Fig. \ref{Fig:Chicken1}(b) with the locations identified in Fig. \ref{Fig:Chicken1}(a). The spectra are qualitatively the same. Differences were identified as a result of the different response of the SVD to raw BCARS spectra versus that of the log-CARS-to-Reference dataset. Retrieving such similarly denoised and processed spectra was $\approx$70$\times$ faster using the fKK-EC methods (average of 3 repeats $\pm$ 1 standard deviation: conventional method $\approx$ 4973 s $\pm$ 26 s total [$\approx$ 7.0 ms / spectrum]; fKK-EC $\approx$ 70 s $\pm$ 3.0 s total [$\approx$ 99 $\mu$s / spectrum]). It should be noted that for the conventional processing, computer memory limitations precluded the processing of the entire image at once; thus, the speed was estimated by performing the KK, PEC, and SEC on 10000 spectra portions of the image and scaling up the time. The SVD/denoising was performed on the whole image. The fKK-EC and ML:fKK-EC were performed on the entire image.

\begin{figure}[htbp]
	\centering\includegraphics[]{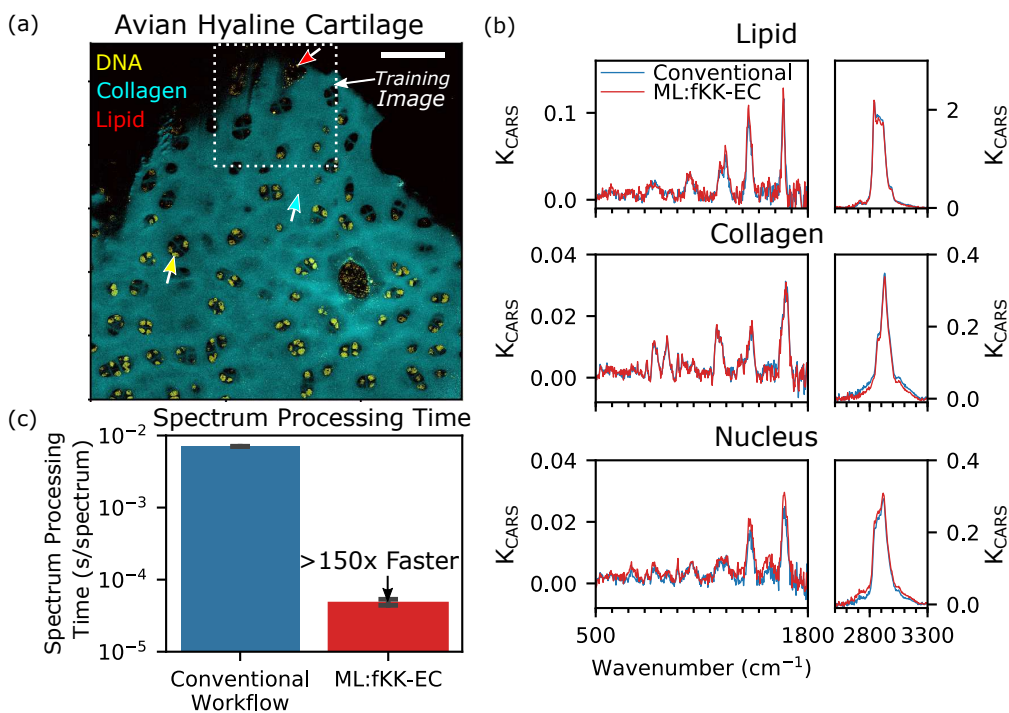}   
	\caption{Pseudocolor image derived from ML:fKK-EC processed CARS image, highlighting DNA (yellow), collagen (cyan), and lipids (red). The dashed-white box indicates the sub-image used for training. Scale bar is 100 $\mu$m. Arrows identify single-pixels used for spectral comparison in (b) between conventional and ML:fKK-EC workflow, which shows close agreement. (b) Single pixel spectra for locations identified by arrows in (a). (c) Comparison of spectrum processing time between conventional and ML:fKK-EC workflow.\label{Fig:Chicken2}}
\end{figure}

Next we processed the same image using the ML:fKK-EC, using 1 of the 9 images as the training image (see Fig. \ref{Fig:Chicken2}(a)). The training image contained 78\thinspace114 spectra. Again the retrieved spectra, see Fig. \ref{Fig:Chicken2}(b), show qualitatively similar results to the conventional workflow with slight noise and baseline differences. Excluding the training time ($<$ 10 s), this method was approximately 150$\times$ faster than the conventional workflow, requiring $<$50 $\mu$s / spectrum to process the entire image. Though these images could have been analyzed in real-time, they were processed after acquisition.

\section{Discussion and conclusion}
Traditionally, the acquisition of CARS spectra was slow, requiring at least tens of milliseconds per spectrum, and most CARS hyperspectral imagery was for a small data size (up to 256 pixels x 256 pixels). Therefore, the speed of individual spectrum-based processing methods was sufficient for the old type of CARS hyperspectral imaging. However, now that the advanced CARS imaging can collect much larger images at a much faster speed, new hyperspectral image processing methods are needed. An additional complication, owing to the inherent distortion of raw CARS spectra, is that the quality and results of an imaging experiment cannot be ascertained until after processing. This, of course, has been a big incentive to use alternative modalities, such as SRS. But as previously described, those alternative modalities do not have the self-referencing ability of CARS, which may be a boon for quantitative analysis. Thus, the aim of this work is the development of high throughput, robust self-referenced Raman signal extraction from CARS spectra with real-time capability.

Though this work demonstrates that the factorization approaches are supremely efficient and capable of being used in a machine learning paradigm, there are still many improvements possible and areas of inquiry for these methods. From a physics/chemistry perspective, we are actively modeling and investigating the nature of the NRB and differences between NRBs of different materials. Further, we are examining the degree to which the real-valued $\chi_{nr}$ assumption is valid in light of multiphoton resonances often found in biomolecules. This information would not only improve quantitative analysis, but as related to this work, it could enable the creation of optimal detrending functions for PEC and SEC (whether factorized version or not). 

There are also many computational lines of inquiry. For example, we are exploring random sampling (``randomized") SVD as a factorization method\cite{Halko2011}, which can approximate the SVD over large datasets orders-of-magnitude faster than traditional SVD. This development could enable real-time processing during all acquisitions (via the ML:fKK-EC) by initially training with few spectra and retraining when it is calculated that the current basis vectors do not adequately support new data. Additionally, we are looking into methods to create a universal basis vector set that could be re-used without training on the current sample. We are also exploring active learning machine learning methods to take advantage of real-time processing that could identify and explore regions of interest during an acquisition.

In conclusion, this work presents the development of a series of new methods for extracting the self-referenced Raman signatures from raw CARS spectra. These new methods, in aggregate, are orders-of-magnitude faster than the conventional implementations and are amenable to high-throughput and even real-time processing with appropriate training data. This advancement facilitates on-the-fly visualization and analysis and would further support such opportunities as \textit{in vivo} imaging and \textit{ad hoc} selection of regions-of-interest.

\section*{Acknowledgments}
The authors wish to thank Sheng Lin-Gibson, Nancy Lin, John Henry Scott, Donald Atha, and Swarnavo Sarkar at NIST for their helpful discussions and insights towards the preparation of this manuscript. 

\section*{Disclosures}
Any mention of commercial products or services is for experimental clarity and does not signify an endorsement or recommendation by the National Institute of Standards and Technology.\\
\\
\noindent The authors declare no conflicts of interest.

\end{document}